\documentclass[iop]{emulateapj}
\usepackage{booktabs}
\usepackage{multirow}
\usepackage{graphicx}
\newcommand{\flux}{{erg~cm$^{-2}$~s$^{-1}$}}
\newcommand{\mflux}{{erg~cm$^{-2}$~s$^{-1}$~Hz$^{-1}$}}
\newcommand{\lum}{{erg~s$^{-1}$}}
\newcommand{\mlum}{{erg~s$^{-1}$~Hz$^{-1}$}}

\usepackage{amsmath}
\setcitestyle{notesep={; }}

\def\simgt{\lower 2pt \hbox{$\, \buildrel {\scriptstyle >}\over {\scriptstyle \sim}\,$}}
\def\simlt{\lower 2pt \hbox{$\, \buildrel {\scriptstyle <}\over {\scriptstyle \sim}\,$}}

\def\chandra{{\it Chandra\/}}

\def\luv{$l_{\mbox{\scriptsize 2500 \AA}}$}

\def\xray{{\hbox{X-ray}}}

\def\aox{$\alpha_{\rm ox}$}
\def\daox{$\Delta\alpha_{\rm ox}$}

\begin{document}
\title{x-ray insights into the nature of quasars with redshifted broad absorption lines}
\author{
Ning-Xiao~Zhang\altaffilmark{1,2}, 
W.~N.~Brandt\altaffilmark{1,2,3},
N.~S.~Ahmed,\altaffilmark{4}
P.~B.~Hall,\altaffilmark{4}
B.~Luo\altaffilmark{5,6},
Scott F. Anderson\altaffilmark{7},
N.~Filiz~Ak\altaffilmark{8,9},
P.~Petitjean\altaffilmark{10},
D.~P.~Schneider\altaffilmark{1,2},
Yue~Shen\altaffilmark{11,12,13},
and R.~Srianand\altaffilmark{14}
}
\altaffiltext{1}{Department of Astronomy and Astrophysics, 525 Davey Lab, The Pennsylvania State University, University Park, PA 16802, USA; npz5018@psu.edu}
\altaffiltext{2}{Institute for Gravitation and the Cosmos, The Pennsylvania State University, University Park, PA 16802, USA}
\altaffiltext{3}{Department of Physics, 104 Davey Laboratory, The Pennsylvania State University, University Park, PA 16802, USA}
\altaffiltext{4}{Department of Physics and Astronomy, York University, Toronto, ON M3J 1P3, Canada}
\altaffiltext{5}{School of Astronomy and Space Science, Nanjing University, Nanjing, Jiangsu 210093, China}
\altaffiltext{6}{Key Laboratory of Modern Astronomy and Astrophysics (Nanjing University), Ministry of Education, Nanjing, Jiangsu 210093, China}
\altaffiltext{7}{Astronomy Department, University of Washington, Box 351580, Seattle, WA 98195, USA}
\altaffiltext{8}{Department of Astronomy and Space Sciences, Faculty of Sciences, Erciyes University, 38039 Kayseri, Turkey}
\altaffiltext{9}{Astronomy and Space Sciences Observatory and Research Center, Erciyes University, 38039 Kayseri, Turkey}
\altaffiltext{10}{Institut d'Astrophysique de Paris, UPMC-CNRS, UMR 7095, 75014 Paris, France}
\altaffiltext{11}{Department of Astronomy, University of Illinois at Urbana-Champaign, Urbana, IL 61801, USA}
\altaffiltext{12}{National Center for Supercomputing Applications, University of Illinois at Urbana-Champaign, Urbana, IL 61801, USA}
\altaffiltext{13}{Alfred P. Sloan Research Fellow}
\altaffiltext{14}{Inter-University Centre for Astronomy and Astrophysics, Post Bag 4, Ganeshkhind, 411\,007, Pune, India}

\begin{abstract}
We present \chandra\ observations of seven broad absorption line (BAL) quasars 
at \hbox{$z=0.863$--2.516} with redshifted BAL troughs (RSBALs). Five of our 
seven targets were detected by \chandra\ in \hbox{4--13~ks} exposures with
ACIS-S. The \aox\ values, \daox\ values, and spectral energy distributions 
of our targets demonstrate they are all \xray\ weak relative to expectations for 
non-BAL quasars, and the degree of \xray\ weakness is consistent with that of 
appropriately-matched BAL quasars generally. Furthermore, our five detected targets show evidence for
hard \xray\ spectral shapes with a stacked effective power-law photon index 
of $\Gamma_{\rm eff}=0.5^{+0.5}_{-0.4}$. These findings support the presence of 
heavy \xray\ absorption ($N_{\rm H}\approx 2 \times 10^{23}$~cm$^{-2}$) in RSBAL quasars, 
likely by the shielding gas found to be common in BAL quasars more generally. 
We use these \xray\ measurements to assess models for the nature of RSBAL
quasars, finding that a rotationally-dominated outflow model is favored 
while an infall model also remains plausible with some stipulations. The
\xray\ data disfavor a binary quasar model for RSBAL quasars in general.
\end{abstract}


\section{INTRODUCTION}

Broad Absorption Lines (BALs) are observed in $\approx 15$\% of optically selected 
quasars within the redshift range of $1.5 \leq z \leq 2.5$ \citep[e.g.,][]{hewett2003,gibson2009}, defined by requiring the velocity width of the BAL absorption trough to be above 2000~km~s$^{-1}$ \citep[e.g.,][]{weymann1991}. 
The intrinsic fraction of BAL quasars, after correcting for observational
selection effects, is even higher \citep[e.g.,][]{hewett2003,dai2008,allen2011}.
The BAL troughs are almost always blueshifted relative to the corresponding emission 
lines in rest-frame ultraviolet (UV) spectra, implying the presence of fast 
outflowing winds. BAL troughs can extend to velocities of at least 
$\approx 60000$~km~s$^{-1}$ \citep[e.g.,][]{rogerson2016}. Outflowing quasar 
winds appear to be a key for understanding how supermassive black holes 
(SMBHs) may be agents of feedback to typical massive galaxies 
\citep[e.g.,][]{chartas2009,fabian2012,arav2013,king2014}.

BAL quasars are commonly classified into one of three groups based on the 
ionization levels of the observed BALs. High-ionization BAL quasars (HiBALs) 
only contain high-ionization BALs such as \ion{C}{4}, \ion{N}{5}, and \ion{O}{6}. 
Low-ionization BAL quasars (LoBALs) show, in addition to the high-ionization
BALs, low-ionization BALs such as \ion{C}{2}, \ion{Al}{3}, and \ion{Mg}{2}. 
Iron low-ionization BAL quasars (FeLoBALs) are LoBALs that also possess 
BALs from \ion{Fe}{2} and/or \ion{Fe}{3}.

BAL quasars usually have low soft \xray\ fluxes compared to 
their optical/UV fluxes \citep[e.g.,][]{green1996,gallagher2006}, and
X-ray spectroscopy reveals that this behavior is often due to heavy and 
complex \xray\ absorption of a nominal-strength underlying \xray\ 
continuum \citep[e.g.,][]{gallagher2002,gallagher2006,fan2009}. 
Thus, the level of \xray\ continuum luminosity, evaluated using
observed \aox\ and \daox, is significantly different between BAL quasars 
and non-BAL quasars. Here \aox\ is defined as 
$0.3838 \log(l_{\rm 2~keV}/l_{2500~\mathring{\rm{A}}})$, 
indicating the relationship between rest-frame \xray\ (2~keV) and 
UV (2500~\AA) luminosity. The quantity \daox\ is 
$\alpha_{\rm ox}({\rm Observed})-\alpha_{\rm ox}(l_{2500~\mathring{\rm{A}}})$, 
representing the observed \aox\ relative to that expected
from the established \hbox{\aox-$l_{2500~\mathring{\rm{A}}}$} relation 
\citep[e.g.,][]{steffen2006}.

\citet{murray1995} proposed an influential accretion-disk wind model for BAL quasars, 
where an equatorial wind is launched from the disk at 
\hbox{$\approx 10^{16}$--$10^{17}$~cm} from the central SMBH 
\hbox{($\approx 10^{8}$--$10^{9} M_{\odot}$)} and radiatively driven 
by UV-line pressure. To accelerate the observed gas to a high velocity
efficiently, this model invokes a ``failed wind'' as shielding 
gas to prevent nuclear \xray\ and extreme-UV (EUV) photons from 
over-ionizing the outflowing gas observed in the UV \citep[e.g.,][]{proga2000}; 
\xray\ absorption by such shielding gas can explain the observed \xray\ 
weakness of many BAL quasars. While this model has had many successes, it 
also faces some challenges. For example, it has been argued that 
the observed level of \xray\ shielding is insufficient to protect the
wind from over-ionization, and that gas clumping may instead be 
responsible for maintaining the needed ionization level 
\citep[e.g.,][]{hamann2013,baskin2014}. Additionally, at least some absorbers are thought to 
be located at kpc-scale distances from the SMBH, leading to 
alternative suggestions about acceleration mechanisms
\citep[e.g.,][]{arav2013,borguet2013}. 

While, as noted above, almost all BALs are blueshifted relative to the 
corresponding emission lines, rare quasars with redshifted BALs (RSBALs) 
have now been identified in significant numbers. Some of these objects have both 
redshifted and blueshifted BALs, while others contain only RSBALs. In the 
large quasar spectroscopic databases of the Sloan Digital 
Sky Survey-I/II/III \citep[SDSS-I/II/III; e.g.,][]{york2000,eisenstein2011}\footnote{Here we refer to spectra taken 
during SDSS-I or SDSS-II as SDSS spectra, and spectra
obtained for the SDSS-III Baryon Oscillation Spectroscopic Survey (BOSS)
as BOSS spectra.}, \citet{hall2013} found 17 BAL quasars with 
RSBALs in \ion{C}{4} and two with \ion{Mg}{2} RSBALs. In this
sample of quasars with RSBALs, the velocity widths of the RSBALs
are above 3000~km~s$^{-1}$, except for one case, and the RSBALs
can extend to redshifted velocities up to about 15000~km~s$^{-1}$. All of the 
three BAL-quasar ionization classes (HiBALs, LoBALs, and FeLoBALs) are 
found for quasars with RSBALs. Notably, the fraction of LoBAL quasars in 
the RSBAL quasar sample is much higher than that in the general population 
of BAL quasars; this result may be a clue to the nature of quasars with RSBALs. 
Among BAL quasars, the objects with RSBALs may provide novel broader
insights about quasar inflows/outflows, and they may represent a new 
method for observing the fueling/feedback of SMBHs.

\citet{hall2013} proposed three models that might explain the nature 
of quasars with RSBALs: a rotationally-dominated outflow model, an 
infall model, and a binary quasar model. These are briefly described
below: 

\begin{enumerate}

\item 
The rotationally-dominated outflow model predicts that redshifted 
absorption can arise when the accretion disk, an extended emission
source, is seen through a rotating outflow launched from the disk. 
This scenario proposes that, at some locations, the outflow has a 
rotational velocity that dominates the component along our line 
of sight of its poloidal outflow velocity
\citep[e.g., see Figure~\ref{fig-rot} and][]{ganguly2001,hall2002}. Such a scenario can 
most likely occur when the accretion disk is viewed at high
inclination.\footnote{The inclination angle represents the angle between the line of sight and the rotational axis of the disk. This definition will be used throughout the paper.} 
This model can naturally explain the presence of
both redshifted and blueshifted absorption when both are 
present, and it can also 
produce only redshifted absorption if the outflow is azimuthally 
asymmetric (so that outflowing material is only present in regions
where the rotational velocity dominates over the component along our line 
of sight of the poloidal outflow velocity). 

\begin{figure}
\includegraphics[trim=0mm 190mm 0mm 170mm, clip, scale=0.10]{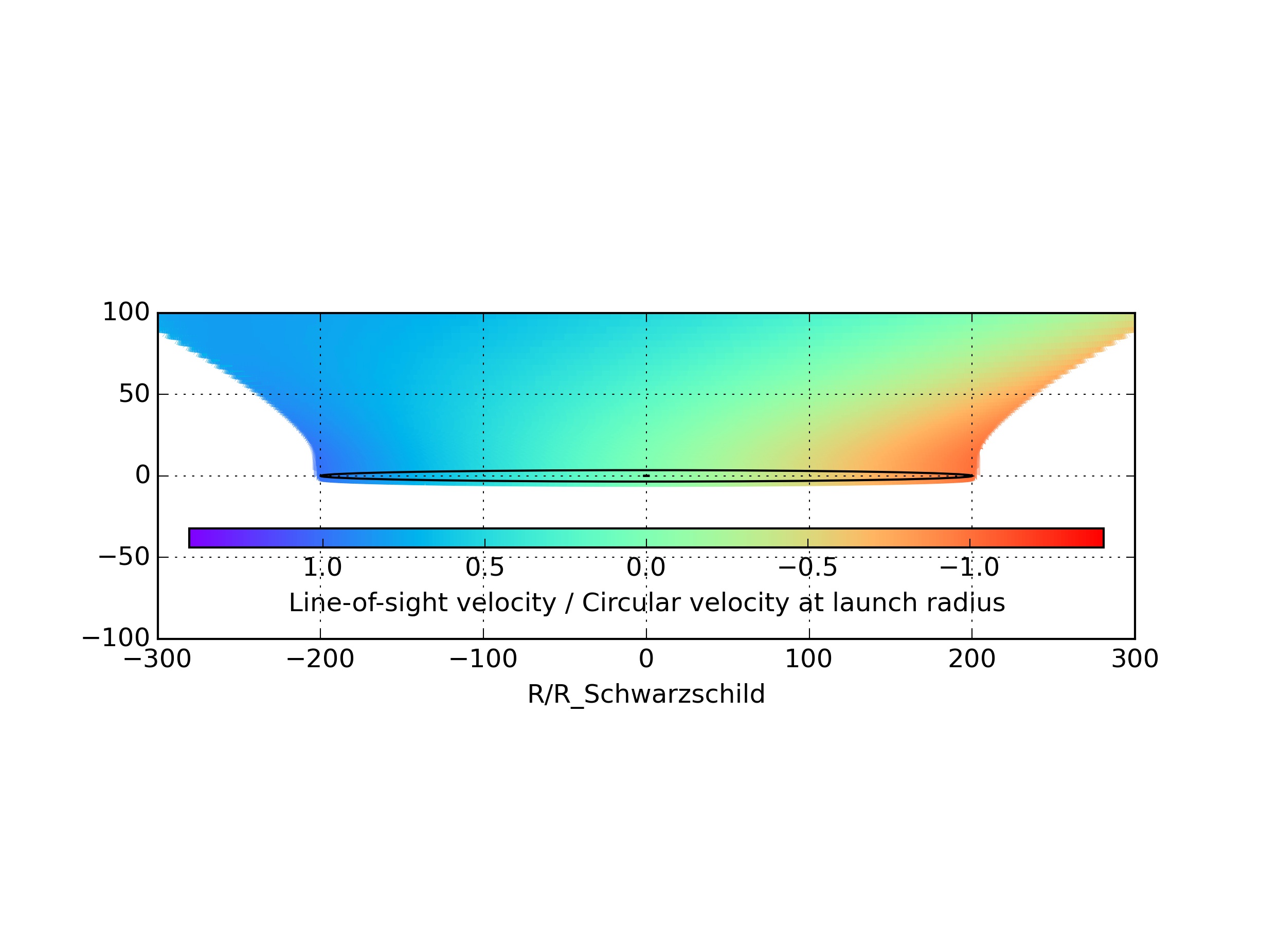}
\includegraphics[trim=0mm 190mm 0mm 170mm, clip, scale=0.10]{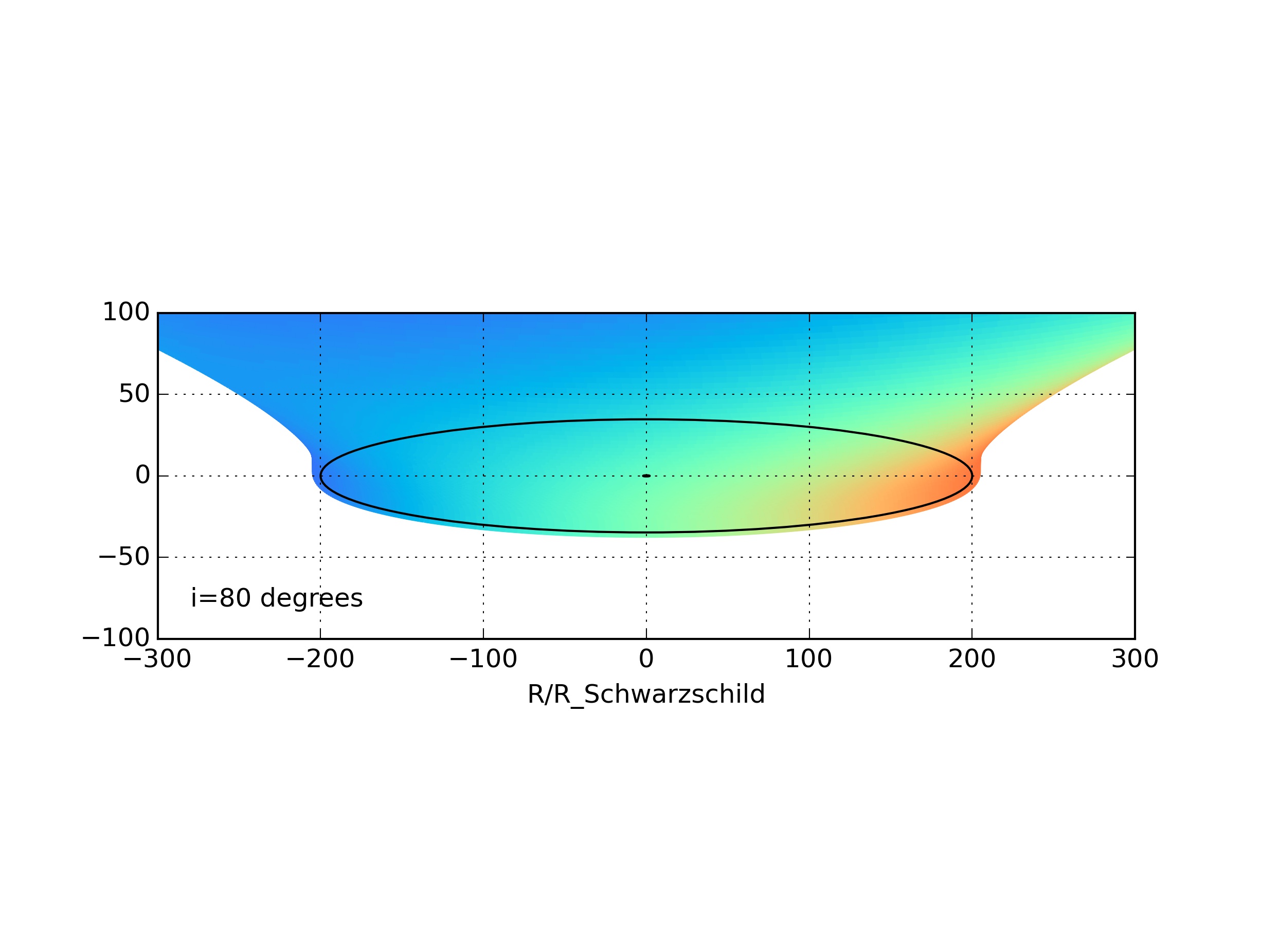}
\includegraphics[trim=0mm 170mm 0mm 170mm, clip, scale=0.10]{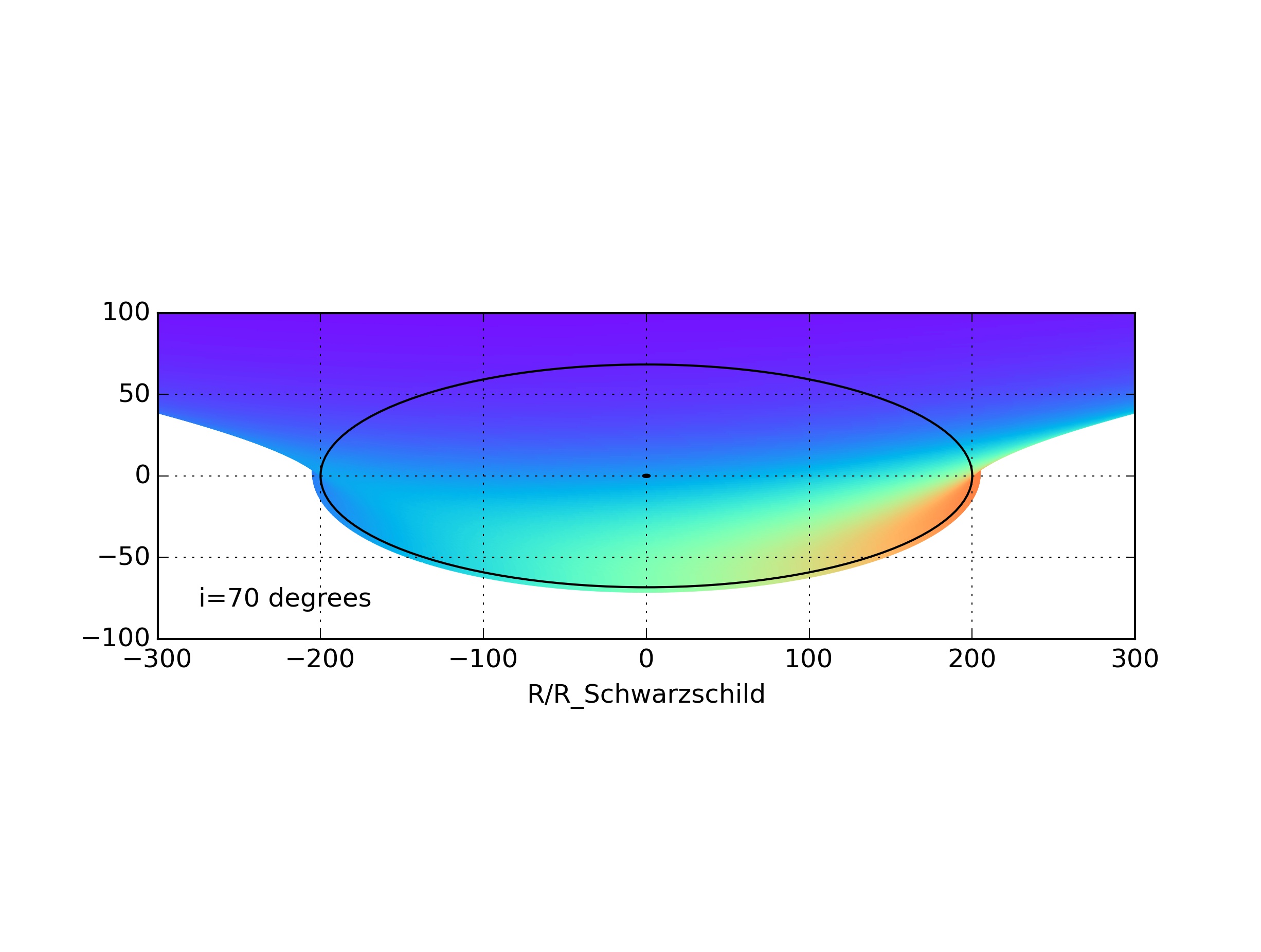}
\caption{Line-of-sight velocity field of the near side of a rotating, accelerating
disk wind, seen at inclination $i=89\arcdeg$ in the top panel and as labelled in 
the other panels.  The central black dot shows the black hole's location.
We see the wind from only one side of the optically thick accretion disk, 
and only the velocities from the near side of the funnel-shaped wind are plotted.
The black ellipse shows the continuum emission region of the disk, 
against which the wind is silhouetted; 
the wind is launched from a narrow annulus just outside that region.
Absorption would only be seen at the velocities found within the black ellipse. 
The ratio of redshifted to blueshifted absorption decreases with decreasing inclination angle.
Different choices for the initial velocity, terminal velocity, acceleration profile, or launching radius  
of the wind would change its velocity field in detail but not qualitatively.
\label{fig-rot}}
\end{figure}

\item  
The infall model proposes that we are observing, via RSBALs, material 
infalling toward the SMBH along the line of sight. To generate 
redshifted absorption extending to the high velocities often observed, 
infall down to a few hundred Schwarzschild radii is required. 
A challenge for this model is that such infalling gas is generally 
expected to have higher ionization levels than observed 
\citep[e.g.,][]{proga2004}. However, infall and disruption of dense
and initially opaque clumps, for which a quasar's radiation pressure
cannot overcome the pull of the SMBH's gravity, might allow gas 
to reach the small required radii while maintaining a low ionization
state.

\item 
The binary quasar model proposes that RSBALs are found in binary quasar 
systems with separations of hundreds of pc to a few kpc
(the kpc-scale upper limit is imposed by current optical-imaging
constraints). In this model, an outflow from the closer, fainter 
member of the binary is backlit by the more distant, brighter
member \citep[e.g.,][]{civano2010,hall2013}. \citet{hall2013} argued 
against the general applicability of the binary model, since basic
estimates of the number of suitable binaries appeared too small 
compared with the observed number of quasars with RSBALs. However, 
further considerations of the timing of the quasar phase in merger 
models, sample-selection effects, and the final-pc problem indicate 
that binary quasars may indeed be sufficiently common that they could 
plausibly explain all quasars with RSBALs 
(E.S. Phinney and P.F. Hopkins 2013, private communication).

\end{enumerate}

\noindent
In this paper, we analyze and interpret exploratory \chandra\ \xray\ 
observations of a sample of seven quasars with RSBALs. The observed targets 
were chosen from the catalog of \citet{hall2013} to have RSBALs in 
either \ion{C}{4} or \ion{Mg}{2}. All targets were selected to have 
bright optical fluxes, allowing sufficiently sensitive \chandra\ 
observations to be obtained economically. We also favored targeting 
objects with no or relatively weak (compared to the RSBAL strength) 
blueshifted BALs; this should help in isolating effects due to RSBALs 
from those due to blueshifted BALs. Aside from defining the basic
\xray\ properties of quasars with RSBALs for the first time, we also 
would like to utilize \xray\ emission to clarify which of the three 
models above best explains the nature of RSBALs. For example, \xray\ 
absorbing shielding gas along the line of sight is expected for the 
rotationally-dominated outflow model, since this model adopts the 
essentials of the standard accretion-disk wind scenario for BAL quasars. 
In contrast, \xray\ absorption is not expected for the binary quasar
model since, in this model, the background quasar producing most of 
the observed \xray\ emission is not launching the
wind that creates the RSBALs. For the infall model, \xray\ absorption
along the line of sight is not automatically expected but is perhaps
possible. 

We describe our sample selection, the utilized \chandra\ observations,
and the \xray\ data analysis in Section~2. To examine the physical 
characteristics of our sample, we present multiwavelength (radio, infrared, 
optical, and UV) data as well as the \xray\ weakness parameter in 
Section~3. In Section~4, we discuss three possible explanations for 
the RSBAL quasars in light of our \xray\ results. Our results are 
summarized in Section~5 where we also discuss future prospects. 
We adopt the cosmological parameters $H_0=67.8$~km~s$^{-1}$~Mpc$^{-1}$, 
$\Omega_{\rm M}=0.308$, and $\Omega_{\Lambda}=0.692$ throughout the paper
\citep[][]{planck2015}.

\section{Sample Selection, Sample Properties, and {\it CHANDRA} Analysis}

\subsection{Sample Selection and Properties}

\begin{figure}
\centerline{
\includegraphics[scale=0.3]{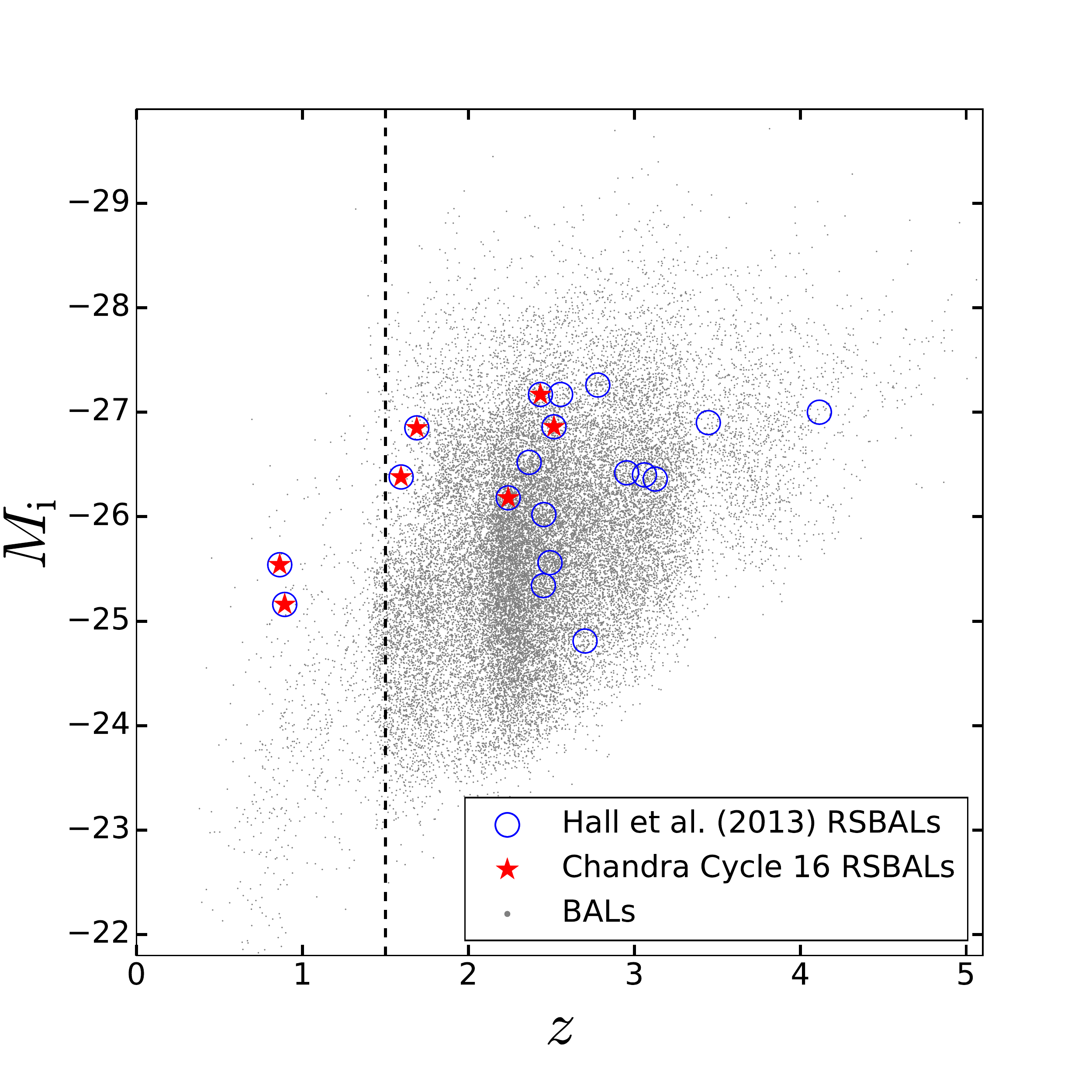}
}
\caption{
The red stars indicate the redshift and SDSS absolute 
$i$-band magnitude of the seven RSBALs in our sample. 
The blue circles show the confirmed RSBALs from \citet{hall2013}.
For comparision, the gray dots represent the 29580 BAL quasars from the 
SDSS DR12 quasar catalog. The vertical dashed line ($z = 1.5$) 
marks the minimum redshift where the spectra of our sample objects start 
to have \ion{C}{4} coverage. The quasars J1125+0029 and 
J1128+0113 are the two RSBALs that lack \ion{C}{4} coverage 
in their observations.
}
\label{fig-mz}
\end{figure}

The redshift and absolute $i$-band magnitude of our targeted RSBAL 
quasar sample compared to the RSBAL quasars in \citet{hall2013} 
are shown in Figure \ref{fig-mz}. Various properties of our targets
 are summarized in Table \ref{table-1}. The RSBAL quasars we targeted with \chandra\ 
have redshifts of \hbox{0.863--2.516}. All seven objects in our \xray\ 
sample were selected from a sample of RSBAL quasars in \citet{hall2013}. 
Among five targeted quasars with redshifted \ion{C}{4} BALs, 
we prioritized those with strong redshifted absorption and weak 
or absent blueshifted absorption (J0830+1654, J1724+3135, and J2157$-$0022); 
this should isolate effects due to redshifted vs. blueshifted 
absorption. For example, the \ion{C}{4} trough in J2157$-$0022 shows a 
sharp edge at small blueshifted velocities (1930~km~s$^{-1}$) and smoothly 
extends to large redshifted velocities of 9050~km~s$^{-1}$. In addition, two 
lower redshift objects, J1125+0029 and J1128+0113 from \citet{hall2013}, 
were selected based on confirmed \ion{Mg}{2} redshifted absorption. Our 
targets were chosen to have bright optical fluxes, with $i$-band apparent 
magnitudes of \hbox{17.9--20.1}, in order to enable suitably sensitive 
\chandra\ observation to be obtained efficiently. To avoid any complicating 
effects associated with jet-linked \xray\ emission \citep[e.g.,][]{miller2011}, 
we required all targets to be radio quiet with radio-loudness 
parameters of $R<10$, where $R=f_{5~{\rm GHz}}/f_{\rm 4400~{\textup{\AA}}}$
\citep{kellermann1989}.

Three BAL groups are included in our sample (see Table \ref{table-1}). 
Six of our targets are LoBALs, as low-ionization absorption is common 
among RSBALs. Two of these six are FeLoBALs. Only one of our 
targets is a HiBAL.

Four of our RSBAL quasars have two epochs of SDSS observations, and all four show
variability of their redshifted \ion{C}{4} or \ion{Mg}{2} absorption 
\citep{hall2013}. Three quasars (J1034+0720, J1628+4744, and J1128+0112 ) only show 
significant variability of their RSBALs, while J1125+0029 shows variability of both its
redshifted and blueshifted BALs. 

\begin{deluxetable*}{lcccccccc}

\tablewidth{0pt}
\tablecaption{RSBAL Quasar Properties}
\tablehead{
\colhead{Object Name}                   	&
\colhead{Redshift}                           	&
\colhead{BAL}					&
\colhead{\ion{C}{4}}             	 	&
\colhead{\ion{C}{4}}              		&
\colhead{\ion{C}{4}}              		&
\colhead{$M_{i}$}                   		&
\colhead{$N_{\rm H,Gal}$}  		&
\colhead{$E(B-V)$}	 			\\
\colhead{}   					&
\colhead{}   					&
\colhead{type}  					&
\colhead{AI$^{\rm tot}$} 			&
\colhead{AI$^{-}$}   				&
\colhead{AI$^{+}$}  				&
\colhead{}   					&
\colhead{}   					&
\colhead{}   					\\
\colhead{(J2000)}   				&
\colhead{}   					&
\colhead{}   					&
\colhead{(km s$^{-1}$)}   			&
\colhead{(km s$^{-1}$)}   			&
\colhead{(km s$^{-1}$)}   			&
\colhead{}   					&
\colhead{($10^{20}$cm$^{-2}$)}   	&
\colhead{}   					\\
\colhead{(1)}         				&
\colhead{(2)}         				&
\colhead{(3)}         				&
\colhead{(4)}         				&
\colhead{(5)}         				&
\colhead{(6)}         				&
\colhead{(7)}         				&
\colhead{(8)}         				&
\colhead{(9)}				
}

\startdata
$083030.26+165444.7$&$2.435$&Lo&2069$\pm$175&743$\pm$72&1326$\pm$102&$-27.17$&$3.8$&0.03\\
$103412.33+072003.6$&$1.689$&Lo\footnote{J1034+0720 has blueshifted low-ionization absorption but no clear redshifted low-ionization absorption.}&2116$\pm$22&2035$\pm$7&81$\pm$28&$-26.85$&$2.9$&0.03\\
$112526.12+002901.3$&$0.863$&FeLo&...&...&...&$-25.54$&$3.7$&0.03\\
$112828.31+011337.9$&$0.893$&FeLo&...&...&...&$-25.16$&$3.3$&0.03\\
$162805.80+474415.6$&$1.595$&Hi&3225$\pm$422&2058$\pm$141&1167$\pm$282&$-26.38$&$1.4$&0.02\\
$172404.44+313539.6$&$2.516$&Lo&3812$\pm$618&0&3812$\pm$618&$-26.86$&$3.3$&0.04\\
$215704.26-002217.7$&$2.240$&Lo&3956$\pm$221&1329$\pm$34&2627$\pm$187&$-26.18$&$5.4$&0.09
\enddata
\tablecomments{
Cols. (1)--(3): Object name, redshift, and BAL type\footnote{``Hi" for high ionization, ``Lo" for low ionization, and ``FeLo" for iron low ionization.} from Table 1 of \citet{hall2013}.
Cols. (4)--(6): The absorption index in km s$^{-1}$ for \ion{C}{4} absorption at all velocities (AI$^{\rm tot}$), \ion{C}{4} blueshifted absorption (AI$^{-}$), and \ion{C}{4} redshifted absorption (AI$^{+}$) from Table 2 of \citet{hall2013}. Two optical spectroscopic observations were available for J1034+0720 and J1628+4744; we selected the one with lower uncertainties.
An entry of ``...'' indicates that \ion{C}{4} is not covered in the optical spectra.
Col. (7): Absolute $i$-band magnitude.
Col. (8): Galactic neutral hydrogen column density in units of $10^{20}$cm$^{-2}$, computed using COLDEN.\footnote{http://cxc.harvard.edu/toolkit/colden.jsp}
Col. (9): The standard Galactic extinction $E(B-V)$ values derived from SDSS extinction values.}
\label{table-1}
\end{deluxetable*}

\subsection{\chandra\ Observations}

Our \chandra\ observations of RSBAL quasars were performed between 2014 Dec~28 and 
2016 April~22 using the Advanced CCD Imaging Spectrometer \citep[ACIS;][]{garmire2003} 
spectroscopic array (ACIS-S). The details of the \chandra\ observations are summarized 
in Table \ref{table-2}. The targets were placed on the
S3 CCD, as is standard. We used VFAINT mode to allow optimal background
removal. The targets have exposure times of \hbox{3.8--13.1~ks}. These 
exposures were set to obtain detections even if our targets are 
\hbox{6--20} times \xray\ weaker than typical quasars, given their optical/UV 
luminosities \citep[e.g.,][]{steffen2006}. This sensitivity level was
required given that BAL quasars are commonly \xray\ weak and/or absorbed 
\citep[e.g.,][]{gallagher2002,gallagher2006,luo2014}. 

\subsection{\xray\ Data Analysis}

We processed the \xray\ data using \chandra\ Interactive Analysis of 
Observations (CIAO) tools. For the data set of each source, we applied 
the {\sc chandra\_repro} script with VFAINT background cleaning. 
Background flares were removed with the {\sc deflare} script using 
sigma clipping at the $3\sigma$ level (little background flaring
was present). The final exposure times are listed in Table~\ref{table-2}. 

To identify our targets, we first ran the {\sc wavdetect} script on the 
soft-band \hbox{(0.5--2.0~keV)}, hard-band \hbox{(2.0--8.0~keV)}, and 
full-band \hbox{(0.5--8.0~keV)} images using the standard wavelet scales 
(i.e., 1.0, 1.414, 2.0, 2.828, and 4.0 pixels) and a false-positive 
probability threshold of $10^{-5}$. Five of the seven targets are 
detected in at least one band within 1.5\arcsec\ of their SDSS 
positions; the two sources not detected by this procedure are 
J0830+1654 and J1128+0113. Next, aperture photometry was performed 
for each target by extracting counts from a circular aperture of 
radius 1.5\arcsec; this aperture size provides a suitable balance 
between capturing source counts and minimizing background counts. 
Background was extracted from an annular region 
with inner radius 10\arcsec\ and outer radius 40\arcsec; background 
point sources in this annulus were removed when measuring the 
background counts. We assessed the significance of the source 
signal by computing a binomial no-source probability, 
$P_{B}$ \citep[e.g.,][]{broos2007,xue2011,luo2013,luo2015}. The definition of $P_{B}$ is
$$P_{B} = \sum^{N}_{X=S} \frac{N!}{X!(N-X)!} p^{X} (1-p)^{N-X}.$$
In this equation, $S$ is the number of source counts; $N$ is the combined 
number of source and background counts; and $p=1$/(1+BACKSCALE), where BACKSCALE 
is the ratio of the areas between the background and the source. In agreement 
with the results from {\sc wavdetect}, five of the seven 
targets were detected in at least one band with a $P_{B}$ value lower than 
0.01 (i.e., a probability of detection above 99\%); the sources not
detected were again J0830+1654 and J1128+0113. The source counts were 
corrected using the enclosed-counts fractions of the \chandra\ point spread 
function (PSF) of 0.951, 0.892, and 0.922 for the soft, hard, and 
full bands, respectively. The net counts for each band, computed from the 
source and background counts, are listed in Table~\ref{table-2}
(with $1\sigma$ uncertainties). For bands where a source is not 
detected, a 90\% confidence-level upper limit is given on the counts
following \citet{kraft1991}.

\begin{deluxetable*}{lccccccccc}
\tablewidth{0pt}
\tablecaption{New \chandra\ Observations and X-ray Photometric Properties of RSBAL quasars}
\tablehead{
\colhead{Object Name}                   	&
\colhead{Observation}                  	&
\colhead{Observation}                  	&
\colhead{Exposure}                   		&
\colhead{Counts}                   		&
\colhead{Counts}                   		&
\colhead{Counts}                   		&
\colhead{Hardness}                   		&
\colhead{$\Gamma_{\rm eff}$}   	&
\colhead{$N_{\textrm H}$}             	\\
\colhead{}   					&
\colhead{ID}   					&
\colhead{Start Date }   			&
\colhead{Time }   				&
\colhead{(0.5--2~keV)}   			&
\colhead{(2--8~keV)}   			&
\colhead{(0.5--8~keV)}   			&
\colhead{Ratio}                   		&
\colhead{}   					&
\colhead{}						\\
\colhead{(J2000)}   				&
\colhead{}   					&
\colhead{(UT)}   				&
\colhead{(ks)}   					&
\colhead{}   					&
\colhead{}   					&
\colhead{}   					&
\colhead{}   					&
\colhead{}   					&
\colhead{($10^{23}$cm$^{-2}$)}   	\\
\colhead{(1)}         				&
\colhead{(2)}         				&
\colhead{(3)}         				&
\colhead{(4)}         				&
\colhead{(5)}         				&
\colhead{(6)}         				&
\colhead{(7)}         				&
\colhead{(8)}         				&
\colhead{(9)}         				&
\colhead{(10)}
}

\startdata
$083030.26+165444.7$&$17043$&2015-05-09&$5.8$&$< 2.4$&$< 2.6$&$<2.5$&...&...&...\\
$103412.33+072003.6$&$17045$&2015-06-29&$4.1$&$< 2.4$&$2.2_{-1.4}^{+2.7}$&$2.1_{-1.4}^{+2.7}$& $> 1.19$& $< 0.8$&$>1.3$\\
$112526.12+002901.3$&$17042$&2015-07-03&$3.8$&$< 2.4$&$2.2_{-1.4}^{+2.7}$&$2.1_{-1.4}^{+2.7}$& $> 1.19$& $< 0.8$&$>0.5$\\
$112828.31+011337.9$&$17046$&2016-04-22&$5.4$&$< 2.4$&$<2.6$&$<2.5$&...&...&...\\
$162805.80+474415.6$&$17044$&2015-08-05&$4.9$&$2.1_{-1.3}^{+2.7}$&$< 2.6$&$2.1_{-1.4}^{+2.7}$&$<1.0$& $> 0.9$&$<0.9$\\
$172404.44+313539.6$&$17040$&2015-11-15&$9.9$&$2.1_{-1.3}^{+2.7}$&$3.3_{-1.7}^{+3.0}$&$5.3^{+3.5}_{-2.3}$&$1.54^{+2.12}_{-0.93}$& $0.5^{+0.9}_{-0.8}$&$4.3^{+5.0}_{-3.3}$\\
$215704.26-002217.7$&$17041$&2014-12-28&$13.1$&$5.2_{-2.2}^{+3.5}$&$6.6_{-2.5}^{+3.7}$&$11.8^{+4.6}_{-3.4}$&$1.28^{+0.95}_{-0.59}$& $0.7^{+0.6}_{-0.5}$&$2.6^{+2.1}_{-1.6}$
\enddata
\tablecomments{
Cols. (1)--(4): Object name, \chandra\ observation ID, observation start date, and background-flare cleaned effective exposure time.
Cols. (5)--(7): Aperture-corrected net counts in the soft (0.5--2~keV), hard (2--8~keV), and full (0.5--8~keV) observed-frame bands. An upper limit at a 90\% confidence level is given if the source is not detected.
Col. (8): Hardness ratio between the hard-band and soft-band counts within a 68\% confidence interval calculated using the BEHR approach.\footnote{http://hea-www.harvard.edu/astrostat/behr/} An entry of ``...'' indicates that
the source is undetected in both bands.
Col. (9): 0.5--8~keV effective power-law photon index, derived from the hardness ratio assuming a power-law spectrum modified by Galactic absorption. An entry of ``...'' indicates that the index cannot be constrained.
Col. (10): The estimated intrinsic neutral hydrogen column densities, derived from the hardness ratios assuming a standard \xray\ power-law spectrum with a photon index of 2.0 (see \S4.1 for further discussion of these quantities).}
\label{table-2}
\end{deluxetable*}

A hardness ratio between hard-band and soft-band counts, along with 
its $1\sigma$ error bar, was computed using the Bayesian Estimation 
of Hardness Ratios (BEHR) approach of \citet{park2006} due to the 
failure of standard error propagation for \xray\ sources with small
numbers of counts (see Table~\ref{table-2}). An effective power-law 
photon index, $\Gamma_{\rm eff}$, was derived from the hardness ratio 
of each source using the Portable, Interactive, Multi-Mission 
Simulator (PIMMS) assuming a power-law spectrum modified by  
Galactic absorption; as expected from the limited numbers of counts, 
these $\Gamma_{\rm eff}$ values for individual sources have
significant uncertainties. Applying stacking of the source counts, 
we also derived a stacked $\Gamma_{\rm eff}$ for the five detected 
targets of $0.5^{+0.5}_{-0.4}$ and a stacked $\Gamma_{\rm eff}$ for 
the four detected LoBAL targets of $0.3^{+0.5}_{-0.4}$. 
We estimate the full-band \xray\ fluxes of our targets from their 
full-band count rates with PIMMSv4.8d using a power-law 
spectrum and $\Gamma_{\rm eff}$ (see Table \ref{table-2}). 
When a source had a lower or upper limit for $\Gamma_{\rm eff}$, 
we used the limit value in this flux calculation. Furthermore, 
based on our stacking results, we adopted $\Gamma_{\rm eff}=0.3$ for 
our two \xray\ undetected LoBAL quasars, J0830+1654 and J1128+0113, 
when deriving upper limits on their full-band fluxes. Our derived 
flux values are not strongly sensitive to the adopted 
$\Gamma_{\rm eff}$. The rest-frame \hbox{2--10~keV} luminosities of our 
RSBAL quasars were computed from their full-band fluxes with the 
standard bandpass correction for redshift.

\section{Multiwavelength Analysis}

\subsection{The X-ray-to-Optical Power-Law Slope}

We list the X-ray-to-optical power-law slopes (i.e., \aox\ values) 
for our sample in Table~\ref{table-3}. 
The observed flux densities at rest-frame 2500~\AA\ for those 
RSBAL quasars (J1034+0720, J1125+0029, J1128+0113, and J1628+4744)
having SDSS spectroscopic observations were 
estimated by normalizing a power-law model with a fixed spectral index
of $-0.5$. To avoid strong absorption and emission lines, 
the initial fitting regions were chosen based on the 
spectral windows described in \citet{gibson2009}. 
For those targets (J0830+1654, J1724+3135, and J2157$-$0022) with only
BOSS observations, we calculated their flux densities at rest-frame 2500~\AA\ 
from the $i$-band\footnote{Since the $i$-band is free from 
strong emission and absorption lines for these three quasars,
its magnitude can better represent the continuum of the spectrum than the $g$-band or $r$-band.
Although rest-frame 2500~\AA\ might be in the $z$-band for those highly redshifted 
quasars, we still utilize the $i$-band magnitude to estimate the flux density at 2500~\AA\ due
to the systematic uncertainties of magnitude in $z$-band whose red end is
defined by the CCDs and not by a filter cutoff.} apparent PSF magnitudes 
in the SDSS DR12 quasar catalog \citep{paris2017} with a K-correction\footnote{We apply a K-correction to transfer the flux density from $i$-band effective 
wavelength to rest-frame 2500~\AA\ assuming a power-law
model with a spectral index of $-0.5$.} and a Galactic-absorption correction.\footnote{The $i$-band PSF magnitude was corrected by the standard 
Galactic extinction listed in Table \ref{table-1}.} The rest-frame 2~keV flux densities were 
calculated from the Galactic absorption-corrected full-band fluxes assuming 
a power-law spectrum with a photon index of $\Gamma_{\rm eff}$ (see \S2.3 for
further details). The \xray\ weakness parameter, $f_{\rm weak}$, 
listed in Table~\ref{table-3} was derived from \daox\ as 
$f_{\rm weak} = 403^{-\Delta \alpha_{\mbox{\tiny OX}}}$, and it
represents the factor by which a quasar is \xray\ weak relative
to the average non-BAL quasar. 

\begin{turnpage}
\begin{deluxetable*}{lcccccccccc}
\tabletypesize{\scriptsize}

\tablewidth{0pt}
\tablecaption{X-ray and Optical Photometric and SED Properties}
\tablehead{
\colhead{Object Name}                   					&
\colhead{Count Rate}                   					&
\colhead{$F_{\rm 0.5-8 keV}$}                   				&
\colhead{$f_{\rm 2~keV}$}                   				&
\colhead{$f_{\rm 2500~\textup{\AA}}$}          			&
\colhead{$\log L_{\rm X}$}                  				&
\colhead{$\log$ \luv}                   					&
\colhead{$\alpha_{\rm OX}$}                   				&
\colhead{$\Delta\alpha_{\rm OX}(\sigma)$}      			&
\colhead{$f_{\rm weak}$}                  					&
\colhead{$R$}  									\\
\colhead{(J2000)}   								&
\colhead{(0.5--8~keV)}   							&
\colhead{($10^{-14}$)}   							&
\colhead{($10^{-33}$)}   							&
\colhead{($10^{-27}$)}   							&
\colhead{(2--10 keV)}   									&
\colhead{}   									&
\colhead{}   									&
\colhead{}   									&
\colhead{}   									&
\colhead{}   									\\
\colhead{}										&
\colhead{($10^{-3}$ s$^{-1}$)}						&
\colhead{(erg cm$^{-2}$ s$^{-1}$)}					&
\multicolumn{2}{c}{(erg cm$^{-2}$ s$^{-1}$Hz$^{-1}$)}	&
\colhead{(erg s$^{-1}$)}							&
\colhead{(erg s$^{-1}$Hz$^{-1}$)}					&
\colhead{}										&
\colhead{}										&
\colhead{}										&
\colhead{}										\\
\colhead{(1)}         								&
\colhead{(2)}         								&
\colhead{(3)}         								&
\colhead{(4)}         								&
\colhead{(5)}         								&
\colhead{(6)}         								&
\colhead{(7)}         								&
\colhead{(8)}         								&
\colhead{(9)}         								&
\colhead{(10)}         								&
\colhead{(11)} 						       		
}

\startdata
$083030.26+165444.7$&$<0.43$ & $<0.84$&$<1.2$&$0.90$&$<43.84$&$31.11$&$<-2.26$&$<-0.61(3.58)$&$>37.8$&$<6.3$\\
$103412.33+072003.6$&$0.53^{+0.67}_{-0.33}$ & $0.79$&$3.2$&$1.67$&$43.75$&$31.10$& $-2.20$&$-0.55(3.23)$&$26.8$&$<2.6$\\
$112526.12+002901.3$&$0.57^{+0.72}_{-0.36}$ &$0.85$&$3.7$&$1.47$&$43.25$&$30.48$& $-2.15$&$-0.59(3.00)$&$34.1$&$<2.2$ \\
$112828.31+011337.9$&$<0.46$&$<0.89$&$<1.9$&$1.02$&$<43.24$&$30.35$&$<-2.20$&$<-0.65(3.42)$&$>50.4$&$<3.1$\\
$162805.80+474415.6$&$0.44^{+0.56}_{-0.28}$ & $0.61$&$2.9$&$1.13$&$43.62$&$30.88$ &$-2.15$&$-0.53(2.93)$&$23.8$&$<4.1$\\
$172404.44+313539.6$&$0.54^{+0.35}_{-0.23}$ &$0.95$&$2.0$&$0.62$&$43.99$&$30.97$ &$-2.11$&$-0.48(2.71)$&$17.4$&$<8.4$\\
$215704.26-002217.7$&$0.90^{+0.35}_{-0.26}$ &$1.44$&$4.6$&$0.42$&$44.18$&$30.72$ &$-1.90$&$-0.31(1.63)$&$6.3$&$<10.0$
\enddata

\tablecomments{
Col. (1): Object name.
Col. (2): Observed 0.5--8~keV \chandra\ count rate in units of 10$^{-3}$~s$^{-1}$.
Col. (3): Galactic absorption-corrected observed-frame 0.5--8~keV flux in units of
$10^{-14}$~\flux, computed using the PIMMSv4.8d\footnote{http://cxc.harvard.edu/toolkit/pimms.jsp} with the power-law photon index in Table 2. When the power-law photon index is an upper or lower limit, we use the limit value for our calculation. We adopt $\Gamma = 0.3$ for two undetected targets J0830+1654 and J1128+0113 (see \S2.3 for details).
Col. (4): Observed flux density at rest-frame 2~keV in units of $10^{-33}$~\mflux, calculated from the unabsorbed 0.5--8~keV flux in the observed frame.
Col. (5): Observed flux density at rest-frame 2500~\AA\ in units of $10^{-27}$~\mflux. Optical SDSS spectroscopic observations were available for J1034+0720, J1125+0029, J1128+0113, and J1628+4744. Based on fitting a power-law model with a fixed spectral index of $-0.5$, we calculated the flux density at rest-frame 2500~\AA\ for each of these SDSS observations. The flux densities of targets (J0803+1654, J1724+3135, and J2157-0022) with only BOSS observations were extrapolated from the SDSS $i$-band apparent PSF magnitudes (see \S3.1 for details).
Col. (6): Logarithm of the rest-frame 2--10~keV luminosity in units
of \lum, derived from the observed 0.5--8~keV flux.
Col. (7): Logarithm of the rest-frame 2500~\AA\ monochromatic luminosity in units
of \mlum.
Col. (8): Measured $\alpha_{\rm OX}$ parameter.
Col. (9): Difference between the measured $\alpha_{\rm OX}$ and the
expected $\alpha_{\rm OX}$ from the \citet{just2007}
\hbox{$\alpha_{\rm OX}$--\luv} relation. 
The statistical significance of this difference,
measured in units of the $\alpha_{\rm OX}$ rms scatter in Table~5 of \citet{steffen2006}, is given in the parentheses.
Col. (10): Factor of X-ray weakness in accordance with $\Delta\alpha_{\rm OX}$.
Col. (11): Radio-loudness parameter, {defined as 
$R=f_{5~{\rm GHz}}/f_{\rm 4400~{\textup{\AA}}}$, where $f_{5~{\rm GHz}}$ and
$f_{\rm 4400~{\textup{\AA}}}$ are the flux densities at
rest-frame 5~GHz and 4400~\AA, respectively.}
}
\label{table-3}
\end{deluxetable*}
\end{turnpage}

\begin{figure*}
\includegraphics[width=\textwidth,angle=0,clip=]{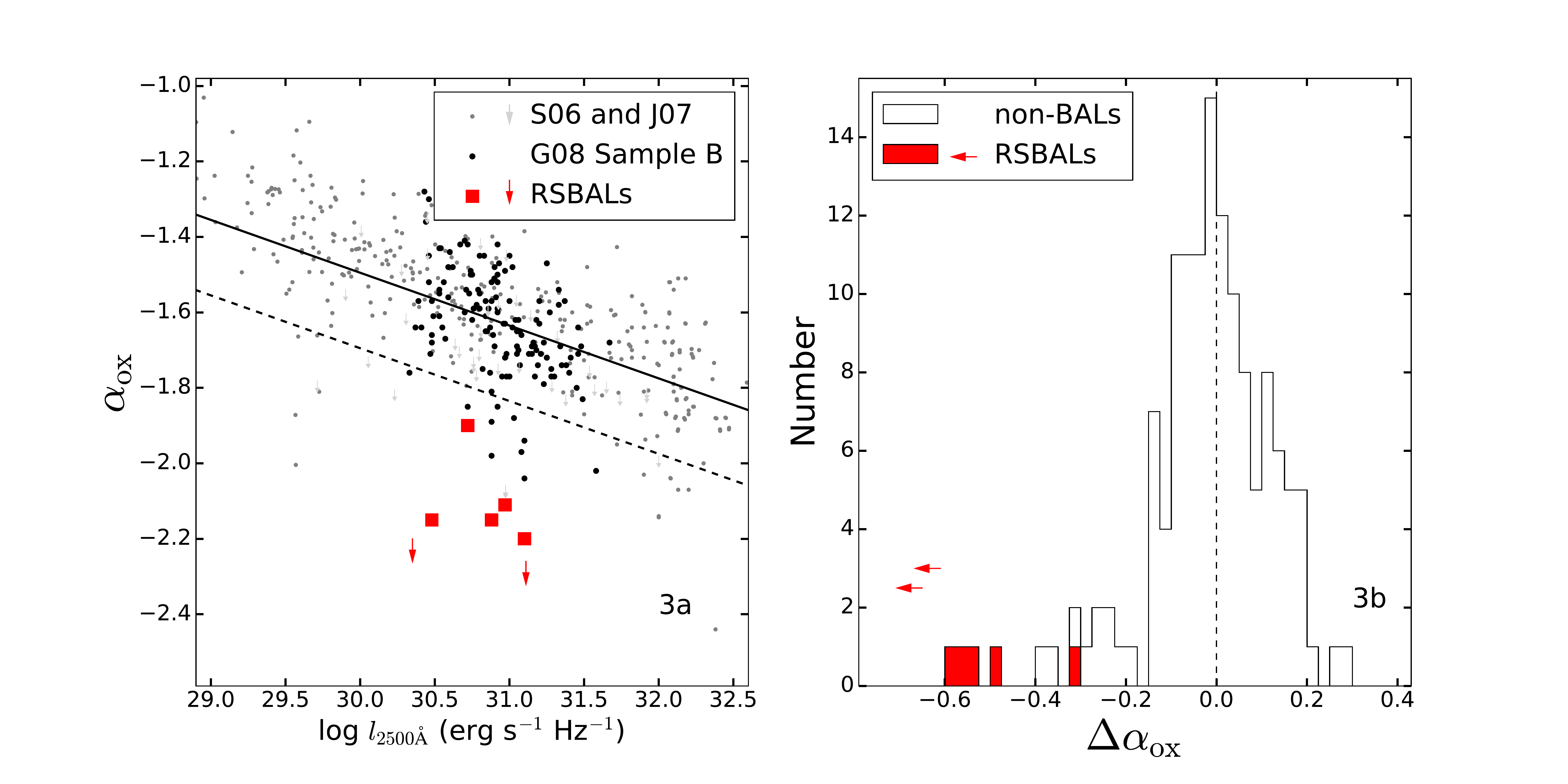}
\caption{Left panel: X-ray-to-optical power-law slope ($\alpha_{\rm ox}$) vs. 
monochromatic luminosity at rest-frame 2500~${\rm \AA}$ 
(\luv) for our RSBAL sample (red squares). 
The RSBAL sample of seven targets was observed by \chandra\ and 
two upper limits (represented by arrows) are shown for J0830+1654 and J1128+0113. Other samples 
in the figure are non-BALs from G08 sample B 
\citep[improved following][]{wu2011}, shown as black dots, and a 
combination of AGN samples from S06 and 
J07 is indicated by gray dots and downward 
arrows. The solid line represents the best-fit relation between 
$\alpha_{\rm ox}$ and \luv\ from \citet{just2007}; 
the dashed line ($\Delta \alpha_{\rm ox} = -0.2$) separates X-ray 
normal and X-ray weak quasars. All seven of our targets are located 
in the X-ray weak region. Right panel: distribution of $\Delta \alpha_{\rm ox}$ for our RSBAL 
sample (shaded histograms) compared with the 132 non-BALs from 
G08 sample B. The two leftward red arrows 
are the upper limits of J0830+1654 and J1128+0113. The 
vertical dashed line indicates $\Delta \alpha_{\rm ox} = 0$. 
All of our targets are in the X-ray-weak wing of the distribution, with six being 
significantly X-ray weaker than any of the G08 Sample B quasars. 
}
\label{fig-xweak}
\end{figure*}

Figure~\ref{fig-xweak}a displays \aox\ vs.\ \luv\ for RSBAL 
quasars. We also show, for comparison purposes, non-BAL quasars 
from \citet[G08]{gibson2008} sample~B \citep[improved following footnote 16 of][]{wu2011}\footnote{
Seven of the sample B quasars are identified as likely BAL quasars in the SDSS DR5 BAL catalog
due to the reconstruction of the \ion{C}{4} emission-line profile or the continuum model. 
In addition, six more sources are identified as BAL or mini-BAL quasars by visual inspections.
These 13 objects have been removed to form the ``improved" sample of \hbox{non-BAL quasars}.} 
and a combination of AGNs from \citet[S06]{steffen2006} and \citet[J07]{just2007}.  
The solid line represents the best-fit relationship between 
\aox\ and \luv\ from \citet{steffen2006}. Following \citet{luo2015}, 
we adopt $\Delta\alpha_{\rm ox}=-0.2$ (the dashed line in Figure~\ref{fig-xweak}a) 
as a reasonable division between \xray\ weak and \xray\ normal quasars. 
All of our targets are located in the \xray\ weak region. In addition, in Figure~\ref{fig-xweak}b
we compare the \daox\ distribution of our sample with that for 
non-BAL quasars from the improved sample~B of G08. 
We adopt the log-rank test \citep[e.g.,][]{feigelson2012}
to assess if our censored RSBAL sample\footnote{
Two objects in our RSBAL sample are not detected in the \xray\ band, hence the censoring.}
follows the same distribution as the 
non-BAL sample. As clearly expected from the visual appearance, 
the resulting $p$-value of 0 demonstrates a significant difference 
between the \daox\ values of RSBAL quasars and non-BAL quasars. 

\begin{figure}
\centerline{
\includegraphics[scale=0.4]{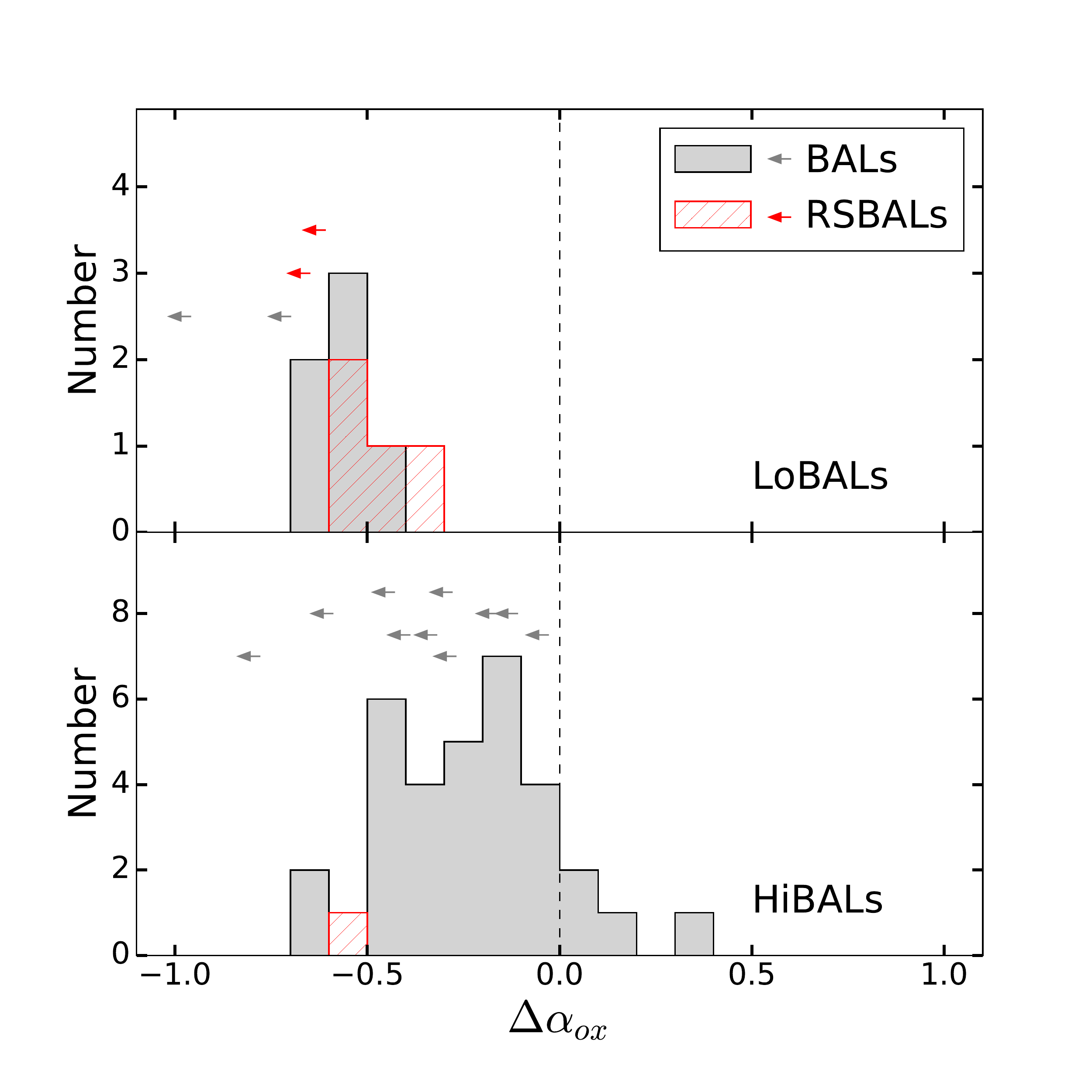}
}
\caption{
Distribution of $\Delta \alpha_{\rm ox}$ for our 
RSBAL sample compared with LoBALs (upper panel) 
and HiBALs (lower panel) from the samples of G09. 
The two red leftward arrows represent upper limits for 
J0830+1654 and J1128+0113. The gray leftward arrows
show upper limits for BAL quasars from the samples of G09.
The vertical dashed line indicates $\Delta \alpha_{\rm ox} = 0$. 
The upper panel shows the LoBAL quasars in our 
RSBAL sample resemble those of typical LoBAL quasars.
In the lower panel, although we only have one HiBAL quasar in our 
RSBAL sample, its \daox value is consistent with 
those HiBAL quasars from the samples of G09.
}
\label{fig-hi_lo_dis}
\end{figure}

We have also compared the \daox\ values of our RSBAL quasars to 
those for LoBAL and HiBAL quasars using the samples from 
\citet[G09;][]{gibson2009} (see Figure~\ref{fig-hi_lo_dis}). 
Comparing the \daox\ values of our low-ionization RSBAL quasars with 
those for LoBAL quasars from G09, we find they appear consistent; the $p$-value of 0.4 
from the log-rank test indicates no statistically significant difference. 
Furthermore, our low-ionization RSBAL quasar sample does show a significant 
difference from the HiBAL quasar sample of G09 ($p$-value $\approx 2 \times 10^{-5}$).
This result demonstrates that the \daox\ values of low-ionization RSBAL
quasars resemble those of typical LoBAL quasars more than those of 
typical HiBAL quasars. While we have only one high-ionization RSBAL quasar, 
its \daox\ value also appears consistent with those of HiBAL quasars from G09.

The \aox\ and \daox\ values of our RSBAL quasars do not seem to
depend obviously upon the relative strengths of the redshifted vs.\
blueshifted UV absorption, although the sample size is too small
for tight constraints in this respect.

Although reddening is commonly observed in the spectra of BAL
quasars and especially in LoBAL quasars \citep[e.g.,][]{trump2006,gibson2009}, 
none of the SDSS spectra of our \chandra\ targets shows strong
reddening such as exhibited by J0941$-$0229 and J1147$-$0250 
in figures 2 and 4, respectively, of \citet{hall2013}.
Thus, we do not
expect that corrections for reddening would materially
change our main results above regarding \aox\ and \daox.
Furthermore, any correction for UV reddening would shift the
\aox\ values for RSBAL quasars toward the lower right in
Figure 3a, where they would remain \xray\ weak.

\subsection{IR-to-X-ray Spectral Energy Distributions}

\begin{figure}
\centerline{
\includegraphics[scale=0.4]{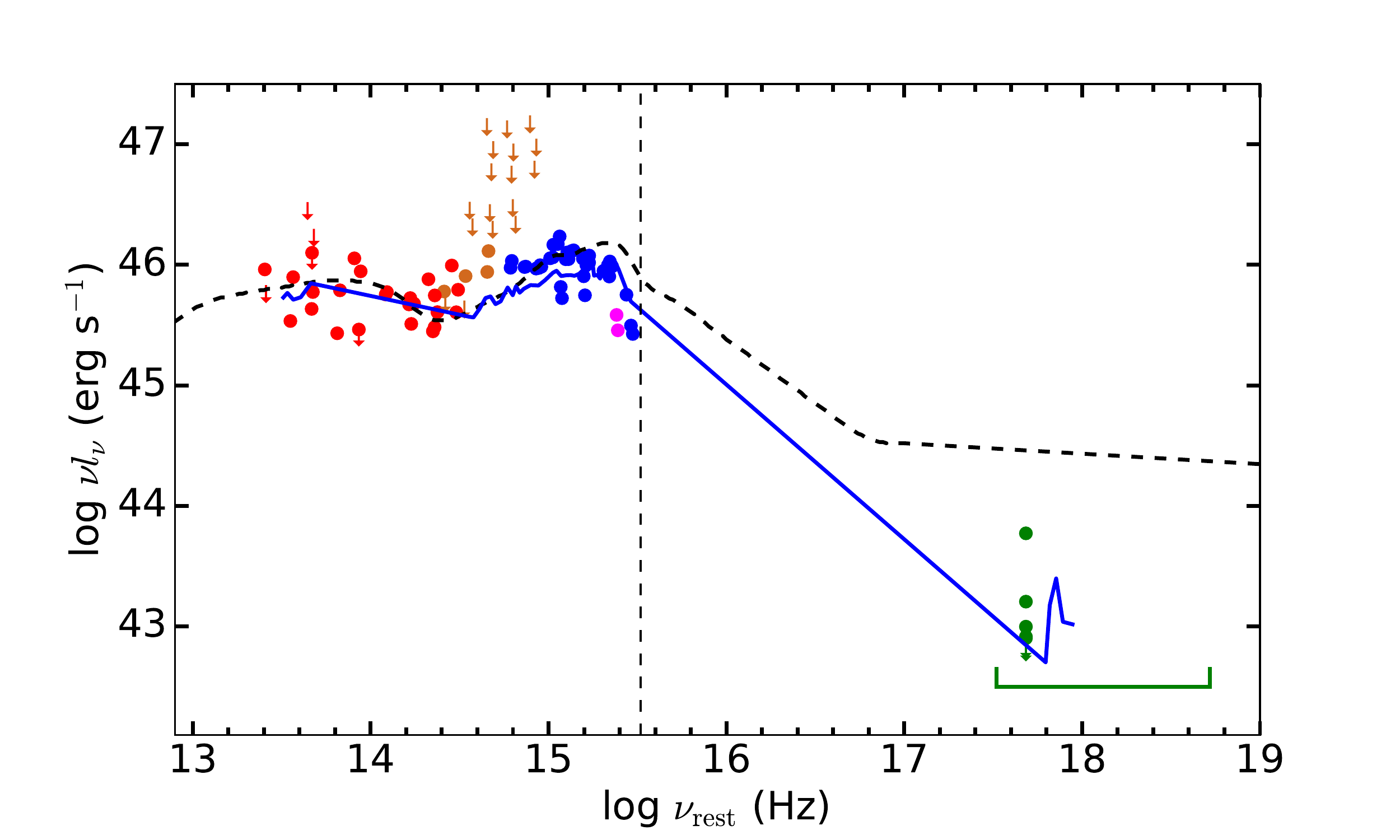}
}
\caption{
Combined SEDs of the seven RSBALs. The IR-to-X-ray SED data are from {\it WISE} (red), 
2MASS (brown), SDSS (blue), {\it GALEX} (magenta), and \chandra\ (green). 
The green segment represents the \chandra\ bandpass in the rest frame
with an average redshift ($z$ = 1.747) of our targets.
The Galactic extinction has been corrected for each band. All of our targets have 
been normalized to the composite SED (dashed line) of optically luminous quasars 
\citep{richards2006} at 3000 \AA\ (corresponding to $10^{15}$~Hz). The data points 
from {\it GALEX} and SDSS {\it u}-band for quasars with redshifts above 0.9 and 2.2, 
respectively, are removed due to the possible presence of a Lyman break feature 
at rest-frame 912~\AA\ (the vertical dashed line at $3.3 \times 10^{15}$~Hz). 
The blue solid line represents the mean SED of BAL quasars from \citet{gallagher2007}.
All of our RSBAL targets are significantly X-ray weak relative to the mean SED of normal
quasars and are consistent with the mean SED of BAL quasars.
}
\label{fig-sed}
\end{figure}

Figure~\ref{fig-sed} presents the suitably normalized infrared-to-\xray\ spectral energy distributions 
(SEDs) of our targets. The data were collected 
from the Wide-field Infrared Survey Explorer ({\it WISE\/}), Two Micron 
ALL Sky Survey (2MASS), Sloan Digital Sky Survey (SDSS), 
{\it Galaxy Evolution Explorer} ({\it GALEX\/}), and \chandra. 

${\it WISE\/}$ observed our targets in four bands centered at 
wavelengths of 3.4, 4.6, 12, and 22~$\mu$m. The 
source flux densities were calculated from {\it WISE\/} Vega 
magnitudes with a color correction \citep{wright2010}.
Three \hbox{high-redshift} RSBAL quasars
(J0830+1654, J1724+3135, and J2157$-$0022) were
not detected at 12 and 22~$\mu$m;
thus we estimated upper limits for these targets.
Two of the seven targets, J1125+0029 and J1128+0113, were 
detected in two of three 2MASS \hbox{near-infrared} bands: $J$ (1.25~$\mu$m), 
$H$ (1.65~$\mu$m), and $K_{\rm s}$ (2.16~$\mu$m). The 
flux densities were calculated from the 2MASS magnitudes referring 
to the fluxes for zero-magnitude from \citet{cohen2003}. For the bands 
with no detection, the 97\% confidence upper limits on flux densities were 
calculated from the 2MASS Atlas Image. For the 
targets without 2MASS detections, we utilized the 
2MASS sensitivity (S/N = 10) to estimate the flux density limit
in each band \citep{skrutskie2006}. For redshifts above 
$\approx 1.4$, the Ly$\alpha$ forest covers the entire 
{\it GALEX\/} NUV (2315.7~\AA) bandpass \citep[e.g.,][]{trammell2007}, 
and thus we only display the {\it GALEX\/} NUV data for the two 
low-redshift targets not significantly affected by Ly$\alpha$ forest 
absorption. The optical and \xray\ calibration were described in 
the previous section. 

We overplot the mean SED of optically luminous SDSS quasars from 
\citet{richards2006} on our data points in Figure~\ref{fig-sed}. 
The SEDs of our targets were scaled to the \citet{richards2006} 
mean SED at rest-frame 3000~\AA\ (corresponding to a frequency of $10^{15}$~Hz). 
The infrared-to-UV SEDs of our targets are broadly consistent 
with the mean SED considering the known \hbox{object-to-object} SED 
scatter for quasars, although there may be a moderate deficit at 
short optical/UV wavelengths. There may also be a moderate
excess at near-infrared wavelengths, but we do not consider this
highly significant owing to the small number of detections and a couple tight upper limits.
To compare with the typical SEDs of BAL quasars, we overplot the mean SED of BAL 
quasars from \citet{gallagher2007}.
As expected from the discussion 
in \S3.1, the \xray\ flux densities at 2~keV are notably low 
relative to the mean SED of normal quasars and
are consistent with the mean SED of BAL quasars.

\subsection{Radio Properties}

The radio-loudness parameters 
\citep[$R = f_{5~{\rm GHz}}$/$f_{\mbox{\scriptsize 4400 \AA}}$;][]{kellermann1989} 
of our seven targets are shown in Table~\ref{table-3}. All of our
targets are radio-quiet quasars with $R<10$. The flux densities at 
rest-frame 5~GHz were computed from the Faint Images of the Radio Sky 
at Twenty-centimeters \citep[FIRST;][]{becker1995} survey at 1.4~GHz 
assuming a radio power-law index of $-0.5$ \citep{kellermann1989}. 
Since our targets are not listed in the FIRST catalog, we estimate 
upper limits for their flux densities at 1.4~GHz as 
\hbox{$0.25+3\sigma$~mJy}, where $\sigma$ is the RMS noise and 
0.25~mJy represents the CLEAN bias \citep{white1997}. The flux densities at 
rest-frame 4400~\AA\ were converted from the flux 
densities at rest-frame of 2500~\AA, using an assumed 
value of the optical power-law index of $-0.5$ \citep{schmidt1983}.

\section{Discussion: X-RAY Assessment of Models For Redshifted BAL Quasars}

Based upon the above multi-wavelength data analyses, all of our RSBAL 
quasars are significantly \xray\ weak. In this section, we will utilize the 
\xray\ results to assess the rotationally-dominated outflow model, the infall 
model, and the binary quasar model; see \S1 for brief details of each model. 
As we will discuss in more detail below, the \xray\ weakness of RSBAL quasars 
can be naturally explained by the rotationally-dominated outflow model, and 
it can also plausibly be described by the infall model with some additional
constraints upon aspects of this model. However, the \xray\ weakness cannot 
be easily explained by the binary quasar model.

\subsection{Rotationally-Dominated Outflow Model}

The rotationally-dominated outflow model adopts the standard accretion-disk 
wind structure commonly used to explain the properties of BAL quasars 
generally \citep[e.g.,][]{murray1995,proga2000}, preferring also
a highly inclined system so that the rotational velocity can dominate the
observed outflow dynamics (see \S1). A key aspect of the standard model is the 
presence of \xray\ absorbing shielding gas located at the base of 
the UV-absorbing wind, likely consisting of optically thick material
that fails to reach escape velocity due to over-ionization. Our \xray\ 
results provide evidence to support the existence of such shielding gas 
in RSBAL quasars.

First, the \xray\ weakness of our RSBAL quasars suggests the existence of 
a heavy \xray\ absorber lying along the line of sight to the small-scale \xray\
emitting region; this \xray\ weakness has been demonstrated using \aox, \daox, 
and examination of SEDs (see \S3). Furthermore, although the number of 
sources is small, the quantitative level of \xray\ weakness for our RSBAL quasars 
is consistent with that for appropriately matched BAL quasars generally 
(see Figure~\ref{fig-hi_lo_dis}). This finding suggests that the \xray\ absorption levels in our 
RSBAL quasars and BAL quasars generally are similar. 

Although the numbers of \xray\ counts for our targets are too limited for
spectral analysis, we have made a basic assessment of likely \xray\ absorption 
levels considering their hardness ratios. If we adopt a standard underlying
\xray\ power-law spectrum with a photon index of $\Gamma=2.0$ (and fixed
Galactic absorption from Table~\ref{table-1}), we can estimate the level of neutral intrinsic 
absorption required to produce the observed hardness ratios (we consider
neutral absorption since this allows straightforward basic comparisons with 
most previous works, although the shielding gas may not be neutral). 
For our four LoBAL quasars with useful hardness-ratio constraints 
(i.e., measurements or lower limits; see Table~2), the estimated neutral 
hydrogen column densities range from $N_{\rm {H}} \gtrsim 5\times 10^{22}$~cm$^{-2}$ 
to $N_{\rm {H}} \approx 4\times 10^{23}$~cm$^{-2}$ (values for each target are listed in  
Table \ref{table-2}).
This wide range is partly due to the redshifts spanned by our targets. Absorption in 
this range is consistent with that often seen for shielding gas in BAL quasars 
generally \citep[$N_{\rm {H}}\approx 10^{22}$--$10^{23.5}$~cm$^{-2}$; 
e.g., ][]{gallagher2002,gallagher2006,fan2009}, again suggesting 
that typical shielding gas is a good candidate for the \xray\ absorber 
in RSBAL quasars. 

In the context of accretion-disk wind models, LoBALs are generally
expected to be observed when our line of sight passes especially
close to the accretion disk \citep[e.g.,][]{murray1995,proga2000,higginbottom2013}. 
There is, moreover, some observational
evidence that this basic idea is correct, at least for many BAL quasars,
based upon correlated variations of ionization levels, kinematics, and
column densities found across BAL-quasar samples
\citep[e.g.,][]{voit1993,filizak2014}. Our \xray\ data are supportive of the 
rotationally-dominated outflow model for RSBAL quasars, and this model
can explain the observed redshifted UV absorption best for large inclinations.
It is exactly for such orientations that LoBALs are most likely to be observed,
potentially explaining in a natural manner the predilection of RSBAL quasars
to show LoBALs in their spectra.

\subsection{Infall Model}

The infall model discussed by \citet{hall2013} suggests the RSBALs 
arise in gas infalling to a few hundred Schwarzschild radii. This
gas would likely need to be in the form of dense clumps in order to 
maintain a sufficiently low ionization level to produce the observed
UV absorption transitions. A high density might arise from compression
during infall by both ram pressure and radiation pressure 
\citep[e.g.,][]{baskin2014}. In the version of this model discussed by 
\citet{hall2013}, one would not necessarily expect shielding gas 
to lie along the line of sight to the \xray\ emitting region. However, 
it could be present in some cases depending upon, e.g., system 
orientation. 

All seven of our targets show evidence for \xray\ absorption at the 
levels expected for appropriately matched BAL quasars generally 
(see \S3 and \S4.1). This finding indicates that, at the least, some 
additional stipulations upon the infall model are likely needed to 
make it agree well with the \xray\ data. For example, one
might reasonably propose that significant small-scale infall can
be observed via UV absorption only in directions where the 
\xray /EUV emission from the quasar is blocked by shielding gas
and does not over-ionize the infalling gas. Gas infalling in other 
directions, or to smaller radii and greater
redshifted velocities, could plausibly end up sufficiently highly
ionized to be observable only at \xray\ wavelengths.  Such gas may have been
 seen to date in at least one AGN \citep[e.g.,][]{giustini2017}.
The observed line of sight in that object appears to show absorption
from redshifted highly ionized iron lines (likely \ion{Fe}{25} and \ion{Fe}{26}).
Modeling of these roughly suggests an ionized absorber column density 
of $\sim 3 \times 10^{23}$ cm$^{-2}$ and redshifted velocity of  
\hbox{$\sim$ 36,000} km~s$^{-1}$. Furthermore,
the majority of RSBAL quasars also possess 
blueshifted BALs in their UV spectra \citep[see Table~1 and][]{hall2013}. 
Given this point, \xray\ shielding gas might indeed be expected along 
the line of sight in the context of the accretion-disk wind model, 
since it is needed to prevent over-ionization of the wind producing 
the blueshifted UV absorption. 

Another possibility is that the \xray\ absorption might arise in 
the same infalling dense clump producing the redshifted UV absorption. 
It is difficult to constrain this model quantitatively at present, 
although it might require somewhat of a coincidence for the 
absorption levels in the infalling clump to match those expected
for the shielding gas. Furthermore, if the \xray\ absorption 
arose in the infalling clump, this configuration would not naturally explain 
why blueshifted UV absorption so commonly accompanies redshifted
absorption. 
 
\subsection{Binary Quasar Model}

In the binary quasar model, we observe a BAL outflow from a 
closer, fainter quasar in the binary that is backlit by 
a more distant, brighter quasar. Thus, the dominant radiation
observed is that from the more distant member of the binary, 
and this member is not generally expected to be 
a BAL quasar (or otherwise heavily \xray\ absorbed). 
The distance between the UV-absorbing material along
the line of sight and the nucleus of the closer, fainter
quasar is typically expected to be much larger than 
the scale of the \xray /EUV absorbing shielding gas
(see \S1), which resides at $\lesssim$ 0.01 pc in the accretion-disk
wind model. One would then not expect any substantial 
\xray\ absorption commonly to be present. Thus, the binary 
quasar model does not provide any natural explanation of the \xray\ 
weakness/absorption found for all seven of our RSBAL quasars, so 
it is disfavored by the \xray\ measurements for the RSBAL quasar
population in general.

\section{Summary and Future Work}

We have presented the \xray\ properties of seven RSBAL quasars 
observed by \chandra, and we have used the results to assess
the available models for these objects. Our main results are
the following: 

\begin{enumerate}

\item
We have compared the \xray\ to optical/UV spectral energy distributions
of our RSBAL quasars with those of non-BAL quasars (from, e.g., S06, J07, G08)
using \aox\ and \daox\ values as well as examination of their SEDs. We
find that all of our RSBAL quasars are notably \xray\ weak compared to non-BAL quasars. 
Furthermore, the quantitative level of \xray\ weakness for RSBAL quasars, ranging from a 
factor of $\approx 6$ to $\simgt 50$, appears similar to that for 
appropriately matched BAL quasars generally. See Sections 3.1 and 3.2. 

\item
The stacked effective power-law photon index derived using the counts
from the five detected RSBAL quasars is $\Gamma_{\rm eff}=0.5^{+0.5}_{-0.4}$. 
This effective photon index is much smaller than that typically found
for radio-quiet quasars ($\Gamma\approx 2$), suggesting the presence of
heavy \xray\ absorption of $N_{\rm H}\approx 2 \times 10^{23}$~cm$^{-2}$ on average. 
This column density is larger than required by the observed UV absorption alone.
However, it is consistent with expectations for the shielding gas of 
accretion-disk wind models as well as measurements of \xray\ absorption in 
BAL quasars generally. See Sections 2.3 and 4.1. 

\item
We have used the \xray\ measurements to assess the available models 
for RSBAL quasars. The \xray\ weakness of RSBAL quasars can be naturally 
explained by the rotationally-dominated outflow model, and this is
our generally favored model. The high system inclinations preferred in
this model may also naturally explain the prevalence of LoBALs in the 
spectra of RSBAL quasars. However, the \xray\ weakness can also 
plausibly be explained by the infall model provided one posits that
\xray\ shielding material always lies along the line of sight where
infall observable in the UV occurs. The \xray\ weakness cannot be easily 
explained by the binary quasar model. See Section~4. 

\end{enumerate}

These \xray\ observations have made an important step toward understanding 
the nature of RSBAL quasars, mainly by demonstrating that they are generally 
\xray\ weak due to absorption by shielding gas (or some other optically thick 
material much like it). However, further work is required to determine 
the precise nature of RSBAL quasars, e.g., by discriminating more strongly
between the rotationally-dominated outflow and the infall models. One 
promising approach involves continued spectroscopic monitoring of the
absorption troughs of RSBAL quasars. As discussed in \S5.3 of 
\citet{hall2013}, the rotationally-dominated outflow model makes the
testable prediction that both redshifted and blueshifted absorption 
troughs should migrate redward as the flow rotates. Such systematic 
migration would not obviously be seen in the infall model, where instead 
one might expect stronger absorption variability at larger redshifted 
velocities \citep[see \S6.1 of][]{hall2013}. Results from such ongoing
spectroscopic monitoring of RSBAL quasars will be presented in 
N. S. Ahmed et al., in preparation. 

Furthermore, additional systematic searches for new RSBAL quasars 
should be performed, now that the sample of SDSS $z$ $>$ 1.5 quasars with 
high-quality spectroscopy has been substantially enlarged relative 
to what was searched in \citet{hall2013}; e.g., see \citet{paris2017}. 
At least 30 new RSBAL quasars should be discovered in such searches, 
and some of these should be suitably bright for further 
efficient \chandra\ observations. 
\\
\\
We thank the referee for helpful feedback.
We also thank S. M.  McGraw for calculating optical flux densities 
by fitting SDSS spectra. We thank C.-T. Chen, 
C.J. Grier, M. Eracleous, E.D. Feigelson, F. Vito, J. Wu, and G. Yang for helpful
discussions. We acknowledge financial support from \chandra\ \xray\ Center 
grant GO5-16092X, NSF grant AST-1516784, and the V.M. Willaman Endowment.
N.S.A. and P.B.H. acknowledge support from NSERC.
B.L. acknowledges support from the National Natural Science Foundation of China
grant 11673010 and the Ministry of Science and Technology of China
grant 2016YFA0400702. N.F.A. acknowledges financial support from 
TUBITAK (115F037). P.P. and R.S. acknowledge support from IFCPR 
under project NO. 5504-B. Y.S. acknowledges support from an Alfred P. Sloan Research 
Fellowship.

\bibliographystyle{apj}
\bibliography{reference.bib}

\end{document}